\documentclass[10pt,amsmath,amssymb,graphicx,nofootinbib,twocolumn]{revtex4} 
\usepackage{bm}
\usepackage{graphics,graphicx,calc,epsfig,pstricks,bbm}
\usepackage{amsmath,amssymb,amsfonts}

\newcommand\ro{\rho}
\newcommand{\be}{\begin{equation}}
\newcommand{\ee}{\end{equation}}
\newcommand{\bea}{\begin{eqnarray}}
\newcommand{\eea}{\end{eqnarray}}

\begin{document}
\title{Purity is not eternal at the Planck scale}

\author{Michele Arzano}
\email{michele.arzano@roma1.infn.it}
\affiliation{Dipartimento di Fisica and INFN,\\ ``Sapienza" University of Rome,\\ P.le A. Moro 2, 00185 Roma, EU}

\begin{abstract}
Theories with Planck-scale deformed symmetries exhibit quantum time evolution in which purity of the density matrix is not preserved. In particular we show that the non-trivial structure of momentum space of these models is reflected in a deformed action of translation generators on operators. Such action in the case of time translation generators leads to a Lindblad-like evolution equation for density matrices when expanded at leading order in the Planckian deformation parameter. This evolution equation is covariant under the deformed realization of Lorentz symmetries characterizing these models. 
\end{abstract}

\maketitle
\noindent

\section{Introduction}
The tension between the basic tenets of quantum theory and black hole quantum evolution \cite{Hawking:1974sw} has been challenging the theoretical physics community for more than thirty years. Two possible scenarios are usually considered for the fate of a radiating black hole: either its quantum evolution is unitary through some yet unknown mechanism or information is lost, pure quantum states can evolve into mixed ones and we must give up unitarity of quantum evolution as we know it when gravitational effects are taken into account.\\
The first option has been by far the most explored in the literature but its viability remains controversial (see e.g. \cite{Susskind:1993if,Kiem:1995iy,'tHooft:1996tq,Almheiri:2012rt}). The possibility of violation of unitary evolution in quantum mechanics has been criticized since its first proposal and deemed as troublesome and leading to conflicts with locality and energy-momentum conservation \cite{Hawking:1976ra,Hawking:1982dj,Banks:1983by}.\\
This letter returns to the issue of the possibility of non-unitary evolution of density matrices and re-examines it in the context of theories with deformed relativistic symmetries. We show, in particular, that a generalized evolution equation of Lindblad type can be realized by time-translation generators belonging to quantum deformations of the Poincar\'e algebra. These models are linked to a certain type of non-commutative field theories in which the coordinate commutators close a non-trivial Lie algebra and in three dimensions are related to the quantization of a relativistic point particle coupled to general relativity. 

\section{Is purity eternal?}
Soon after the discovery of black hole quantum radiance, Hawking pointed out that a full evaporation of these objects could lead to the evolution of pure quantum states into mixed states \cite{Hawking:1976ra}. Indeed the vacuum state of a freely falling observer appears to a fiducial observer outside the black hole, who traces out the degrees of freedom inside the horizon, as a thermal state at the Hawking temperature. The mixed nature of this state is due to the existence of the unobservable region inside the horizon for the observer outside the black hole and to the entangled nature of the vacuum state of the quantum field. Once the black hole has completely evaporated there are no longer degrees of freedom inside the horizon and thus we are left with a mixed state. Having started from a pure vacuum state, albeit perceived by the observer at infinity at any time as a mixed one due to her inability to observe the black hole interior, we are left with a mixed state at the end of the evaporation thus violating unitarity of quantum evolution.\\
Banks, Peskin and Susskind \cite{Banks:1983by} put under scrutiny the proposal by Hawking of a fundamental non-unitary evolution in quantum gravity \cite{Hawking:1982dj}. Looking for the most general evolution equation preserving hermiticity and trace of the density matrix they re-discovered the {\it Lindblad equation} \cite{Lindblad:1975ef}
\be\label{Lind1}
\dot{\rho}=-i [H, \rho] -\frac{1}{2}h_{\alpha\beta}\left(Q^{\alpha}Q^{\beta}\rho+\rho\, Q^{\beta} Q^{\alpha}- 2 Q^{\alpha} \rho\, Q^{\beta}\right)
\ee
where $h_{\alpha\beta}$ is a hermitian matrix and $Q^{\alpha}$ form a basis of hermitian matrices with $Q^{0}=1$. As shown in \cite{Banks:1983by} the evolution equation (\ref{Lind1}) preserves positivity of $\rho$ if the matrix $h_{\alpha\beta}$ is {\it positive-definite} and it leads to increasing entropy in the evolution if $h_{\alpha\beta}$ is {\it real}. In addition time evolution {\it does not} lead to violations of energy conservation if $h_{\alpha\beta}$ is real and positive and the operators $Q^{\alpha}$ commute with the Hamiltonian (uniquely defined as the operator appearing in the commutator in the right hand side of (\ref{Lind1})). Global conserved charges commute with the Hamiltonian but they are non-local (integrals of local currents) and evolution according to (\ref{Lind1}) was shown in this case to lead to non-local effects which manifest as violations of ``cluster decomposition" \cite{Banks:1983by} . Almost ten years later Srednicki \cite{Srednicki:1992bp} re-considered the problem and actually showed that such non-locality is quite harmless and {\it does not} lead to observably macroscopic violations of locality. Using global charges and requiring {\it Lorentz covariance} of the Lindblad equation, however, he showed that for $h$ real time evolution eventually leads to a non-positive definite density matrix. This caused \cite{Srednicki:1992bp} to reject the possibility of a Lorentz covariant Lindblad equation with $Q^{\alpha}$ given by global charges associated to space-time symmetries. The additional conceptual drawback with Lindblad time evolution raised in \cite{Srednicki:1992bp} concerned the fact that there was no evidence of whether  (\ref{Lind1}) had ``any reasonable chance to arise as the low energy limit of a more fundamental theory."\\
Below we will show that the framework of quantum Poincar\'e algebras allows for deformed time translation generators whose adjoint action leads to a covariant Lindblad equation. In three space-time dimensions we will show that gravity itself dictates such deformation of relativistic symmetries. We will further show how a similar scenario for symmetry deformation is realized in four dimensions for certain models linked to non-commutative space-times.

\section{Deforming translations}
A fundamental principle of quantum mechanics states that symmetry generators act on tensor product states describing composite systems according to the Leibnitz rule and on observables via the commutator i.e. the {\it adjoint action}. These are basic concepts of the theory of representation of a Lie algebra on a Hilbert space and an algebra of operators, respectively, which admit generalizations.\\
Consider for definiteness a free scalar field on Minkowski space. A generic element of the Poincar\'e algebra $\mathfrak{p}\equiv\mathfrak{iso}(3,1)$ will act on the one-particle Hilbert space $\mathcal{H}$ which by definition carries an irreducible representation of $\mathfrak{p}$. In the standard textbook treatments of canonical quantization one considers a basis of $\mathcal{H}$ given by eigenstates of the translation generators
\be
P_{\mu} | k \rangle = k_{\mu}| k \rangle\,.
\ee
This representation extends to the dual space $\mathcal{H}^*$ spanned by bras 
\be
P_{\mu} \langle k | = - k_{\mu} \langle k |\,,
\ee
while the action ``from the right" is analogous to the ``action to the left" on kets i.e. $\langle k | P_{\mu} = k_{\mu} \langle k |$.
Multi-particle states are described by symmetrized tensor products of the one-particle space $\mathcal{H}$ and the representation of $P_{\mu}$ on a generic Fock space element is given by a generalized Leibniz rule encoded in the ``second quantization" of $P_{\mu}$ 
\begin{multline}
P_{\mu}+ (P_{\mu} \otimes \mathbbm{1} +\mathbbm{1}\otimes P_{\mu})+\\
 + (P_{\mu}\otimes \mathbbm{1}\otimes \mathbbm{1} + \mathbbm{1}\otimes P_{\mu} \otimes \mathbbm{1}+ \ \mathbbm{1}\otimes \mathbbm{1}\otimes P_{\mu})+...
\end{multline}
where the first term acts on one-particle states the second on two-particle states and so on. The notions above can be used to derive the
action of symmetry generators on observables/operators. Indeed it suffices to derive the action for the projection operator $\pi_{k}= | k \rangle \langle k |$ seen as the ``outer product" of a ket and a bra namely $\pi_{k}= \pi (| k \rangle \otimes \langle k |) = | k \rangle \langle k |$ the action is defined by
\begin{multline}\label{adj1}
P_{\mu}(\pi_{k}) =  \pi (P_{\mu}(| k \rangle \otimes \langle k |)) = \\
= \pi ( P_{\mu}| k \rangle \otimes \langle k | -  | k \rangle \otimes \langle k | P_{\mu}) = [P_{\mu}, \pi_{k}]
\end{multline}
which is nothing but the known {\it adjoint action} of symmetry generators on operators.\\
These very basic notions of representation theory admit an interesting generalization. Consider a basis of kets $| k(g) \rangle$ for the one-particle Hilbert space labelled by {\it coordinates on a non-abelian Lie group} rather than on Minkowski space. Since these kets should describe the states of a quantum relativistic particle a sensible requirement is that the Lorentz group acts {\it transitively} on the momentum group manifold. A set of commuting generators of translations associated with this basis of kets will act as an ordinary representation i.e.
\be
P_{\mu} | k(g) \rangle = k_{\mu}(g)| k(g) \rangle
\ee
However when acting on the dual representation the non-trivial structure of momentum space must be taken into account and the action of translation generators is given by
\be
P_{\mu} \langle k(g)| = k_{\mu}(g^{-1})\langle k(g) |\,.
\ee
This action is dictated by the definition of dual representation and by the group homomorphism properties of coordinate functions. Indeed in order to comply with the homomorphism requirement momentum eigenvalues $k_{\mu}(g)$ must follow a non-abelian composition rule 
\be
k_{\mu}(g)\oplus k_{\mu}(h)\equiv k_{\mu}(gh)\neq k_{\mu}(h)\oplus k_{\mu}(g)\equiv k_{\mu}(hg)\,,
\ee
and 
\be
k_{\mu}(g)\oplus k_{\mu}(g^{-1})= k_{\mu}(gg^{-1}) = k_{\mu}(\mathbbm{1}) =0\, .
\ee
This might look just like an odd realization of translation generators on a Hilbert space but it actually reflects a well defined notion of representation of deformed symmetries according to the theory of Hopf algebras. Models with this kind of symmetries have been studied for their connection with non-commutative field theories in four and three space-time dimensions \cite{Kosinski:1999ix,AmelinoCamelia:2001fd,Agostini:2006nc,Arzano:2007gr,Freidel:2007hk,Arzano:2010jw,Sasai:2007me,Arzano:2013sta} and in the latter case are directly related to the quantization of a relativistic point particle coupled to Einstein gravity \cite{deSousaGerbert:1990yp,'tHooft:1996uc,Meusburger:2005mg,Schroers:2007ey}. As an illustrative example I will briefly discuss this simpler three-dimensional model below and then move on to a more realistic four dimensional case.

\section{Deformed symmetries and Lindblad evolution in $2+1$ gravity}
General relativity in three space-time dimension has the well known feature of not possessing local degrees of freedom \cite{Staruszkiewicz:1963zza}. Relativistic point particles can be coupled to the theory as point-like defects whose world-line spans a string. The resulting geometry is a conical space-time characterized by a deficit angle $\alpha=8\pi G m$ proportional to the particle's mass $m$ where $G$ is the three-dimensional Newton's constant \cite{Deser:1983tn}. The momentum of such particle/defect is 
given by an element $h \in SL(2,\mathbb{R})$ such that \cite{Matschull:1997du}
\be\label{conjcond}
\frac{1}{2}\mathrm{Tr}(h) = \cos(4\pi G m)\, .
\ee
The subspace of $SL(2,\mathbb{R})$ determined by the condition (\ref{conjcond}) should be thought of as the analogous of the mass-shell for a relativistic point particle in three dimensions. Indeed it is easily shown that in the limit $G\rightarrow 0$ equation (\ref{conjcond}) ``flattens" to the ordinary mass-shell condition \cite{Arzano:2014ppa}.\\
In analogy with ordinary relativistic quantum theory, a basis of kets $| k(g) \rangle$ labelled by elements of $SL(2,\mathbb{R})$ constrained by (\ref{conjcond}) will span the Hilbert space of a quantum relativistic particle of given mass \cite{Arzano:2013sta}. These kets will be eigenstates of a maximal commuting set of {\it deformed} translation generators $P_{\mu}$ with a non-trivial action on the dual Hilbert space and on tensor products.  Without going into the mathematical details (which the interested reader can find e.g. in \cite{Koornwinder:1996uq,Bais:1998yn,Bais:2002ye,Schroers:2014dua}) it suffices here to say that $P_{\mu}$ are the generators of the translation sector of a ``quantum deformation" of the Poincar\'e group known as the {\it quantum double} of the Lorentz group. To appreciate the non-trivial structure introduced by the deformation let us first look at the properties of the eigenvalues of the kets $| k(g) \rangle$. Let us introduce the group parametrization
\be
g=\sqrt{1+(4\pi G)^2 k^2}\, \mathbb{I} + 4\pi G\, k_{\mu} \gamma^{\mu}
\ee
where $\gamma^{\mu}$ are two-by-two traceless matrices and $\mathbb{I}$ is the identity matrix. This assigns coordinates $k_{\mu}(g)=1/2\pi G\, \mathrm{Tr}(g\gamma_{\mu})$ to a group element. The non-trivial composition of momenta, at leading order in $G$, is given by
\be\label{comp1}
k_{\mu}(gh)=k_{\mu}(g)+ k_{\mu}(h)+ 4\pi G \epsilon_{\mu\nu\sigma} k^{\nu}(g) k^{\sigma}(h)
\ee
where $ \epsilon_{\mu\nu\sigma}$ is the three dimensional Levi-Civita symbol, while coordinates for the inverse group element remain ``undeformed" 
\be
k_{\mu}(g^{-1})= -k_{\mu}(g)\, .
\ee
It is suggestive to notice that the non-trivial term in (\ref{comp1}) can be considered as a Planck-scale deformation of the usual composition of momenta since in three dimensions the  Newton's constant $G$ is proportional to the inverse Planck mass. In the limit of small momenta the deformation simply vanishes. Borrowing the terminology and notation from Hopf algebra theory and introducing a Planckian mass scale $\kappa=1/4\pi G$ we say that the relation (\ref{comp1}) above reflects a non-trivial {\it co-product} for translation generators
\be\label{coprod3}
\Delta P_{\mu} = P_{\mu} \otimes \mathbbm{1} + \mathbbm{1}\otimes P_{\mu} + \frac{1}{\kappa}\, \epsilon_{\mu\nu\sigma} P^{\nu} \otimes P^{\sigma}\,,
\ee
meaning that the generators $P_{\mu}$ act on tensor product representations according to a deformed Leibniz rule, while their {\it antipodes}, i.e. their actions on dual representations, are undeformed
\be
S(P_{\mu})= -P_{\mu}\, .
\ee
The action of $P_{\mu}$ on operators is determined by the {\it quantum adjoint action} \cite{Ruegg:1994bk,Arzano:2007nx}, a generalization of the adjoint action (\ref{adj1}) to the case of symmetry generators with non-trivial co-product and antipode. Specializing to a density matrix $\rho$ this reads
\be\label{qadjact}
\mathrm{ad}_{P_{\mu}} (\ro) =(\mathrm{id}\otimes S)\Delta(P_{\mu}) \diamond \ro
\ee
where $(a\otimes b) \diamond c \equiv a\, c \,b $. This can be seen as a generalization of the action (\ref{adj1}) for deformed translation generators which in the limit of vanishing deformation parameter (i.e. trivial $\Delta P_{\mu}$ and $S(P_{\mu})$) reproduces the usual {\it action by commutator} or adjoint action. For the time translation generator $P_0$ we thus have the deformed adjoint action
\be
\mathrm{ad}_{P_0} (\ro) = [P_0, \ro] - \frac{1}{\kappa} \, \epsilon_{0ij} P^i \ro\, P^j\, .
\ee
leading to a {\it deformed evolution equation}
\be 
\dot{\rho}= -i [P_0, \ro] + \frac{i}{\kappa} \, \epsilon_{0ij} P^i \ro\, P^j\, ,
\ee
where the time derivative in the left hand side should intended with respect to the time translation parameter associated to a finite time translation generated by $P_0$. This evolution equation can be rewritten in Lindblad form as
\be\label{lind3}
\dot{\rho}=-i [P_0, \rho] -\frac{1}{2}h_{ij}\left(P^i P^j \rho+\rho\, P^j P^i - 2 P^j \rho\, P^i\right)\, .
\ee
where the ``dissipation" matrix is given by
\be\label{dmatrix}
h=
\frac{i}{\kappa}\begin{pmatrix}
0&0&0\\
0&0&1\\
0&-1&0\\
\end{pmatrix}\,.
\ee
This deformed evolution equation is a {\it leading order} result derived from the leading order deformation of the co-product (\ref{coprod3}). Of course a result at all orders in $1/\kappa$ can be easily obtained from using (\ref{qadjact}) and the known expression of the deformed co-product $\Delta P_0$. The resulting deformed evolution equation will no longer look like an ordinary commutator plus correction terms but like a ``non-linear" commutator and thus it will not be of the Lindblad form. Notice also that equation (\ref{lind3}) is Lorentz covariant in a ordinary sense. Indeed for the type of deformation encoded into the quantum double construction it turns out that the Lorentz sector is left untouched and the action of Lorentz transformations on momenta can be easily obtained from the action of $SL(2,\mathbb{R})$ on itself by conjugation \cite{Bais:1998yn,Sasai:2007me,Arzano:2014ppa}. The commutators describing the adjoint action of boosts and rotations on the generators $P_{\mu}$ close the ordinary Poincar\'e algebra.\\
Thus this symmetry deformation scenario reproduces at leading order a covariant Lindblad equation involving global charges associated to translation symmetry. The ``dissipation" matrix (\ref{dmatrix}) is not positive definite therefore positivity of the evolution of $\rho$ is not guaranteed. In \cite{Srednicki:1992bp} it was argued, using a specific example, that for real $h$ this Lindblad evolution scheme will eventually lead to a non-positive definite $\rho(t)$. In the case of (\ref{lind3}) the dissipation matrix $h$ {\it is not} real and it is by no means obvious if a similar argument can be carried through. A thorough analysis of this aspect is beyond the scope of the present Letter but it will certainly be a primary issue in future investigations.

\section{Four dimensional Lindblad evolution and deformed symmetries} 
The simplest way to generalize the picture above to four space-time dimensions is to take as momentum space a four dimensional non-abelian Lie group. In the most studied example this is achieved by considering the Iwasawa decomposition \cite{Barut:1986dd,KowalskiGlikman:2004tz} of the de Sitter group $SO(4,1)$ into the Lorentz group $SO(3,1)$ and the {\it abelian nilpotent} group $AN(3)$. Denoting with $J^{\mu\nu}\,,\,\,\,(\mu\,,\nu=0,..,4)$ the generators belonging to the de Sitter algebra $\mathfrak{so(4,1)}$ the group $AN(3)$ is obtained by exponentiating the four-dimensional Lie sub-algebra given by the four generators $J^{4\mu}+J^{0\mu}$. These can be written in the following five dimensional matrix representation and introducing a constant $\kappa$ with dimensions of energy as
\begin{align}
X_0 = -\frac{i}{\kappa} \left(
\begin{array}{ccc}
0\! & {\mathbf 0}\! & 1 \\ 
{\mathbf 0}\! & {\mathbf 0}\! & {\mathbf 0} \\ 
1\! & {\mathbf 0}\! & 0 
\end{array}
\right),\,\, 
X_a = \frac{i}{\kappa} \left(
\begin{array}{ccc}
0\! & {\mathbf e}^t_a\! & 0 \\ 
{\mathbf e}_a\! & {\mathbf 0}\! & {\mathbf e}_a \\ 
0\! & -{\mathbf e}^t_a\! & 0 
\end{array}
\right),
\end{align}
where ${\mathbf e}_a$ is a basis of unit three vectors and ${\mathbf e}_a^t$ are their transpose. Such generators satisfy the commutation relations $[X_0, X_a] = i/\kappa X_a$, $[X_a, X_b] = 0$ namely those of $\kappa$-Minkowski {\it non-commutative space-time} \cite{Majid:1994cy}. We write a generic $AN(3)$ element as $g = e^{i p^a X_a} e^{i p_0 X_0}$ and introducing coordinates 
\begin{align}
k_0 & = \kappa \sinh\left(\tfrac{p_0}{\kappa}\right) + \frac{1}{2\kappa} e^{p_0/\kappa} p_ap^a\,, \nonumber\\ 
k_a & = -e^{p_0/\kappa} p_a\,, \nonumber\\ 
k_4 & = \kappa \cosh\left(\tfrac{p_0}{\kappa}\right) - \frac{1}{2\kappa} e^{p_0/\kappa} p_ap^a\,,
\end{align}
it is easily checked that 
\be
-k_0^2+\vec{k}^2+k_4^2=\kappa^2
\ee
with $k_0+k_4>0$. Thus the $AN(3)$ momentum space manifold is half of four dimensional de Sitter space. The constant $\kappa$ determines its curvature and it is usually viewed as a Planckian energy scale, in analogy with the three dimensional example discussed before.\\
Translation generators associated with the group parametrization in terms of $k_{\mu}$ exhibit non-trivial co-products and antipodes reflecting non trivial composition of momenta and their inversion. Their expression for time translation generators, at leading order in the (inverse) Planckian parameter $\kappa$ is given by \cite{Borowiec:2009vb}
\bea
\Delta(P_0) & = & P_0 \otimes  \mathbbm{1} + \mathbbm{1} \otimes P_0 +\frac{1}{\kappa} P_m \otimes P_m\,,\label{coprk}\\
S(P_0) & = & -P_0 +\frac{1}{\kappa} \vec{P}^2\,. \label{antik}
\eea
The fact that we have a non-trivial antipode for $P_0$ indicates that the notion of dual representation will be deformed and thus that of the conjugate action on operators as we will see in the following. Another key difference with the $2+1$-dimensional case is the non-trivial role played by Lorentz boosts. It turns out that when the Lorentz sector is included in the picture consistency relations between the action of $SO(3,1)$ on $AN(3)$ and vice-versa \cite{Majid:1994cy,Majid:2008iz} impose that Lorentz boosts also exhibit non-trivial co-products and antipodes
\bea
\Delta(N_i) & = & N_i \otimes \mathbbm{1} +  \mathbbm{1} \otimes N_i - \frac{1}{\kappa} P_0 \otimes N_i - \frac{1}{\kappa} \epsilon_{ijm} P_i \otimes M_m\,\nonumber \\
S(N_i) & = & - N_i \left(1+ P_0/\kappa \right)- \frac{1}{\kappa} \epsilon_{ijm} P_j M_m\,.
\eea
The resulting picture is that of a quantum deformation of the Poincar\'e algebra known as $\kappa$-Poincar\'e algebra \cite{Lukierski:1992dt,Lukierski:1991pn}. The deformed adjoint action of the time translation generator, obtained form the leading order expressions (\ref{coprk}) and (\ref{antik}), on a density matrix leads in this case to a {\it non-symmetric} Lindblad equation
\be\label{lind4}
\dot{\rho}= -i [P_0, \ro] + \frac{i}{\kappa}P_m \rho P_m - \frac{i}{\kappa} \rho\, \vec{P}^2\,.
\ee
From a comparison with (\ref{Lind1}) and (\ref{lind3}) one would indeed expect that in order to preserve hermiticity of $\rho$ an extra $ \vec{P}^2 \rho$ term in the equation above should have been present. However, as mentioned before, a non-trivial notion of antipode for $P_0$ also implies a deformed notion of hermitian adjoint for the quantum adjoint action. Indeed in the presence of a deformed antipode $S(P_0)\neq -P_0$ the hermitian adjoint will be now given by $(\mathrm{ad}_{P_0}(\cdot))^{\dagger}\equiv \mathrm{ad}_{S(P_0)}(\cdot)$ so that quantum adjoint action satisfies the the following {\it deformed} skew-hermiticity condition $(\mathrm{ad}_{P_0}(\cdot))^{\dagger}\circ \mathrm{ad}_{P_0}(\cdot)=0$.\\
Let us also stress the peculiar role of Lorentz boosts in this context. While the momentum eigenvalues $k_{\mu}$ transform as ordinary Lorentz four-vectors and the adjoint action of boosts and rotations on the translation generators $P_{\mu}$ closes an undeformed Poincar\'e algebra, the transformation properties of the modified Lindblad equation (\ref{lind3}) 
are non trivial. Indeed the quantum adjoint action of boosts on the density matrix (and on a generic operator) is deformed according to
\be
\mathrm{ad}_{N_i}(\rho) = [N_i,\rho] + \frac{1}{\kappa} [P_0,\rho] N_i + \frac{1}{\kappa} \epsilon^{ijm} [P_j,\rho] M_m\,.
\ee
However it can be verified that the quantum adjoint actions of $N_i$ and $P_0$ satisfy 
\be
\mathrm{ad}_{N_i}(\mathrm{ad}_{P_0}) (\cdot) - \mathrm{ad}_{P_0}(\mathrm{ad}_{N_i}) (\cdot) = \mathrm{ad}_{\mathrm{ad} N_i(P_0)}  (\cdot)\, .
\ee
which ensures that the evolution equation (\ref{lind3}) obeys a deformed notion of Lorentz covariance according to the non-trivial quantum adjoint action.

\section{Discussion}
The results presented illustrate that a Lindblad-like evolution equation is a natural consequences in theories where time translation generators are deformed within the framework of quantum algebras. This observation opens two main fronts on which further investigations must focus. On one side it will be important to establish whether the Planck-scale Lindblad equations derived here signal a departure from unitarity or they reflect a {\it deformed} notion of unitarity to be understood in the context of Hopf algebras. The resulting picture will be conceivably relevant in addressing basic questions in the description of black hole quantum evolution. On the other hand it should be noted that there exist a long history of experimental searches of departures from ordinary quantum time evolution of the Lindblad type (see e.g. \cite{Ellis:1983jz,Adler:1995xy,Ellis:1995xd,AmelinoCamelia:2010me}). Thus the possibility of non-unitary evolution in models of deformed relativistic symmetries could open a new phenomenological window for testing these scenarios of Planck scale physics besides the avenues already explored in the literature \cite{AmelinoCamelia:1997gz, AmelinoCamelia:2000ge,AmelinoCamelia:2000zs,Ng:1999hm,Gambini:2006yj,Albert:2007qk,Hossenfelder:2010zj,AmelinoCamelia:2008qg} .  

\vspace{-.5cm}
\begin{acknowledgments}
\vspace{-.5cm}
I would like to thank G. Gubitosi and J. Kowalski-Glikman for discussions. This research was supported by a Marie Curie Career Integration Grants within the 7th European Community Framework Programme and in part by a grant from the John Templeton Foundation.  
\end{acknowledgments}


\begin{thebibliography}{99}

\bibitem{Hawking:1974sw} 
  S.~W.~Hawking,
  Commun.\ Math.\ Phys.\  {\bf 43}, 199 (1975)
  [Erratum-ibid.\  {\bf 46}, 206 (1976)].

\bibitem{Susskind:1993if} 
  L.~Susskind, L.~Thorlacius and J.~Uglum,
  Phys.\ Rev.\ D {\bf 48}, 3743 (1993)
  [hep-th/9306069].

\bibitem{Kiem:1995iy} 
  Y.~Kiem, H.~L.~Verlinde and E.~P.~Verlinde,
  Phys.\ Rev.\ D {\bf 52}, 7053 (1995)
  [hep-th/9502074].

\bibitem{'tHooft:1996tq} 
  G.~'t Hooft,
  Int.\ J.\ Mod.\ Phys.\ A {\bf 11}, 4623 (1996)
  [gr-qc/9607022].

\bibitem{Almheiri:2012rt} 
  A.~Almheiri, D.~Marolf, J.~Polchinski and J.~Sully,
  JHEP {\bf 1302}, 062 (2013)
  [arXiv:1207.3123 [hep-th]]; S. L. Braunstein, 
[arXiv:0907.1190 [quant-ph]] published as S. L. Braunstein, S.
Pirandola and K. Zyczkowski, 
Phys.\ Rev.\ Lett. 110, 101301 (2013) 


\bibitem{Hawking:1976ra} 
  S.~W.~Hawking,
  Phys.\ Rev.\ D {\bf 14}, 2460 (1976).

\bibitem{Hawking:1982dj} 
  S.~W.~Hawking,
  Commun.\ Math.\ Phys.\  {\bf 87}, 395 (1982).

\bibitem{Banks:1983by} 
  T.~Banks, L.~Susskind and M.~E.~Peskin,
  Nucl.\ Phys.\ B {\bf 244}, 125 (1984).

\bibitem{Srednicki:1992bp} 
  M.~Srednicki,
  Nucl.\ Phys.\ B {\bf 410}, 143 (1993)
  [hep-th/9206056].

\bibitem{Lindblad:1975ef} 
  G.~Lindblad,
  Commun.\ Math.\ Phys.\  {\bf 48}, 119 (1976).


\bibitem{Kosinski:1999ix} 
  P.~Kosinski, J.~Lukierski and P.~Maslanka,
  Phys.\ Rev.\ D {\bf 62}, 025004 (2000)
  [hep-th/9902037].  
  
\bibitem{AmelinoCamelia:2001fd} 
  G.~Amelino-Camelia and M.~Arzano,
  Phys.\ Rev.\ D {\bf 65}, 084044 (2002)
  [hep-th/0105120].  
  
\bibitem{Agostini:2006nc} 
  A.~Agostini, G.~Amelino-Camelia, M.~Arzano, A.~Marciano and R.~A.~Tacchi,
  Mod.\ Phys.\ Lett.\ A {\bf 22}, 1779 (2007)
  [hep-th/0607221]. 
  
\bibitem{Arzano:2007gr} 
  M.~Arzano and A.~Marciano,
  Phys.\ Rev.\ D {\bf 75}, 081701 (2007)
  [hep-th/0701268];  
  Phys.\ Rev.\ D {\bf 76}, 125005 (2007)
  [arXiv:0707.1329 [hep-th]].   
  
\bibitem{Freidel:2007hk} 
  L.~Freidel, J.~Kowalski-Glikman and S.~Nowak,
  Int.\ J.\ Mod.\ Phys.\ A {\bf 23}, 2687 (2008)
  [arXiv:0706.3658 [hep-th]].  
  
\bibitem{Arzano:2010jw} 
  M.~Arzano,
  Phys.\ Rev.\ D {\bf 83}, 025025 (2011)
  [arXiv:1009.1097 [hep-th]].
  
\bibitem{Sasai:2007me} 
  Y.~Sasai and N.~Sasakura,
  Prog.\ Theor.\ Phys.\  {\bf 118}, 785 (2007)
  [arXiv:0704.0822 [hep-th]]. 
 
  
\bibitem{Arzano:2013sta} 
  M.~Arzano, J.~Kowalski-Glikman and T.~Trzesniewski,
  Class. Quantum Grav. {\bf 31} (2014) 035013
  [arXiv:1305.6220 [hep-th]].  


\bibitem{deSousaGerbert:1990yp}
  P.~de Sousa Gerbert,
  Nucl.\ Phys.\ B {\bf 346} (1990) 440.

\bibitem{'tHooft:1996uc} 
  G.~'t Hooft,
  Class.\ Quant.\ Grav.\  {\bf 13}, 1023 (1996)
  [gr-qc/9601014].

\bibitem{Meusburger:2005mg}
  C.~Meusburger and B.~J.~Schroers,
  Nucl.\ Phys.\ B {\bf 738} (2006) 425
  [hep-th/0505143].
    
\bibitem{Schroers:2007ey}
  B.~J.~Schroers,
  PoS QG {\bf -PH}, 035 (2007)
  [arXiv:0710.5844 [gr-qc]].
  
\bibitem{Staruszkiewicz:1963zza}
  A.~Staruszkiewicz,
  Acta Phys.\ Polon.\  {\bf 24}, 735 (1963).

\bibitem{Deser:1983tn}
  S.~Deser, R.~Jackiw and G.~'t Hooft,
  Annals Phys.\  {\bf 152}, 220 (1984).

\bibitem{Matschull:1997du}
  H.~J.~Matschull and M.~Welling,
  Class.\ Quant.\ Grav.\  {\bf 15}, 2981 (1998)
  [arXiv:gr-qc/9708054].

\bibitem{Arzano:2014ppa} 
  M.~Arzano, D.~Latini and M.~Lotito,
  arXiv:1403.3038 [gr-qc].  



\bibitem{Koornwinder:1996uq} 
  T.~H.~Koornwinder and N.~M.~Muller,
  q-alg/9605044.

  
\bibitem{Bais:1998yn} 
  F.~A.~Bais and N.~M.~Muller,
  Nucl.\ Phys.\ B {\bf 530}, 349 (1998)
  [hep-th/9804130].  

\bibitem{Bais:2002ye} 
  F.~A.~Bais, N.~M.~Muller and B.~J.~Schroers,
  Nucl.\ Phys.\ B {\bf 640}, 3 (2002)
  [hep-th/0205021].

\bibitem{Schroers:2014dua} 
  B.~Schroers and M.~Wilhelm,
  arXiv:1402.7039 [hep-th].
  
  
\bibitem{Ruegg:1994bk} 
  H.~Ruegg and V.~N.~Tolstoi,
  Lett.\ Math.\ Phys.\  {\bf 32}, 85 (1994)
  [hep-th/9406146].  
  
\bibitem{Arzano:2007nx} 
  M.~Arzano,
  Phys.\ Rev.\ D {\bf 77}, 025013 (2008)
  [arXiv:0710.1083 [hep-th]].  
  
\bibitem{Barut:1986dd} 
  A.~O.~Barut and R.~Raczka,
  Singapore, Singapore: World Scientific ( 1986) 717p 
  
\bibitem{KowalskiGlikman:2004tz} 
  J.~Kowalski-Glikman and S.~Nowak,
  hep-th/0411154. 
  
\bibitem{Borowiec:2009vb} 
  A.~Borowiec and A.~Pachol,
  J.\ Phys.\ A {\bf 43}, 045203 (2010)
  [arXiv:0903.5251 [hep-th]].  
  
\bibitem{Majid:1994cy} 
  S.~Majid and H.~Ruegg,
  Phys.\ Lett.\ B {\bf 334}, 348 (1994)
  [hep-th/9405107].  
  
\bibitem{Majid:2008iz} 
  S.~Majid and B.~J.~Schroers,
  J.\ Phys.\ A {\bf 42}, 425402 (2009)
  [arXiv:0806.2587 [gr-qc]].  
  
\bibitem{Lukierski:1992dt} 
  J.~Lukierski, A.~Nowicki and H.~Ruegg,
  Phys.\ Lett.\ B {\bf 293}, 344 (1992)
  
\bibitem{Lukierski:1991pn} 
  J.~Lukierski, H.~Ruegg, A.~Nowicki and V.~N.~Tolstoi,
  Phys.\ Lett.\ B {\bf 264}, 331 (1991).
  
\bibitem{Ellis:1983jz} 
  J.~R.~Ellis, J.~S.~Hagelin, D.~V.~Nanopoulos and M.~Srednicki,
  Nucl.\ Phys.\ B {\bf 241}, 381 (1984).  

\bibitem{Adler:1995xy} 
  R.~Adler {\it et al.}  [CPLEAR Collaboration],
  Phys.\ Lett.\ B {\bf 364}, 239 (1995)
  [hep-ex/9511001].

\bibitem{Ellis:1995xd} 
  J.~R.~Ellis, J.~L.~Lopez, N.~E.~Mavromatos and D.~V.~Nanopoulos,
  Phys.\ Rev.\ D {\bf 53}, 3846 (1996)
  [hep-ph/9505340].  
  
\bibitem{AmelinoCamelia:2010me} 
  G.~Amelino-Camelia  {\it et al.},
  Eur.\ Phys.\ J.\ C {\bf 68}, 619 (2010)
  [arXiv:1003.3868 [hep-ex]].  
  
\bibitem{AmelinoCamelia:1997gz} 
  G.~Amelino-Camelia, J.~R.~Ellis, N.~E.~Mavromatos, D.~V.~Nanopoulos and S.~Sarkar,
  Nature {\bf 393}, 763 (1998)
  [astro-ph/9712103].

\bibitem{AmelinoCamelia:2000ge} 
  G.~Amelino-Camelia,
  Int.\ J.\ Mod.\ Phys.\ D {\bf 11}, 35 (2002)
  [gr-qc/0012051].
 
\bibitem{AmelinoCamelia:2000zs} 
  G.~Amelino-Camelia and T.~Piran,
  Phys.\ Rev.\ D {\bf 64}, 036005 (2001)
  [astro-ph/0008107].
  
\bibitem{Ng:1999hm} 
  Y.~J.~Ng and H.~van Dam,
  Found.\ Phys.\  {\bf 30}, 795 (2000)
  [gr-qc/9906003].  

\bibitem{Gambini:2006yj} 
  R.~Gambini, R.~Porto and J.~Pullin,
  Gen.\ Rel.\ Grav.\  {\bf 39}, 1143 (2007)
  [gr-qc/0603090].

\bibitem{Albert:2007qk} 
  J.~Albert {\it et al.}  [MAGIC and Other Contributors Collaborations],
  Phys.\ Lett.\ B {\bf 668}, 253 (2008)
  [arXiv:0708.2889 [astro-ph]].

\bibitem{Hossenfelder:2010zj} 
  S.~Hossenfelder,
  arXiv:1010.3420 [gr-qc].

\bibitem{AmelinoCamelia:2008qg} 
  G.~Amelino-Camelia,
  Living Rev.\ Rel.\  {\bf 16}, 5 (2013)
  [arXiv:0806.0339 [gr-qc]].  

  
\end{thebibliography}
\end{document}